\documentclass{aa}
\input epsf

\begin{document}

\thesaurus{1(10.7.2; 10.01.1; 10:07:3 \object{2MASS GC01}; 
	10:07:3 \object{2MASS GC02}; 08.08.1; 08.04.1)} 

\title{Deep Near Infrared Photometry of New Galactic Globular Clusters
  \thanks{Based on data obtained at the ESO-NTT in La Silla, Chile.}}

\author{Valentin D. Ivanov\inst{1} 
  \and Jordanka Borissova\inst{2} 
  \and Leonardo Vanzi\inst{3}}

\offprints{Valentin D. Ivanov}

\institute{Steward Observatory, The University of Arizona, Tucson, 
	AZ 85721, U.S.A., vdivanov@as.arizona.edu
  \and Institute of Astronomy, Bulgarian Academy of Sciences and 
	Isaac Newton Institute of Chile, Bulgarian Branch, 
	72 Tsarigradsko $\rm chausse\grave{e},$ 1784 Sofia, Bulgaria, 
	jura@haemimont.bg
  \and European Southern Observatory (ESO), Alonso de Cordova 3107, 
	Santiago - Chile}

\date{Received 24 08 2000 / Accepted 13 09 2000}
\maketitle

\begin{abstract}
We present a preliminary report on the first deep infrared photometry 
of \object{2MASS GC01} and \object{2MASS GC02} (GC01 and GC02 hereafter) - 
new Galactic globular cluster (GC) candidates, discovered by the 2MASS. 
They both appear to be genuine disk GCs though highly obscured. The 
location of the two GCs suggests that they are metal rich - 
[Fe/H]$\sim-$0.5. We estimated their distance and reddening using 
the $\rm K_s-band$ brightness of the red giant branch (RGB) bump, and 
$\rm J-K_s$ color of the RGB at $\rm M_K=-3$ mag: $\rm D=3.1\pm0.5$, 
$3.9\pm0.6$ kpc, and $\rm A_V=20.9\pm0.7$, $17.2\pm1.2$ mag, for 
\object{GC01} and \object{GC02} respectively. Variation in the metal 
abundance can change our results within $30-35$\%.
\end{abstract}

\keywords{The Galaxy: globular clusters: general --
        The Galaxy: abundances --
	The Galaxy: globular clusters: individual: \object{2MASS GC01} -- 
	The Galaxy: globular clusters: individual: \object{2MASS GC02} -- 
	Stars: Hertzsprung-Russel and C-M diagrams -- 
	Stars: distances}

\section{Introduction}
The known Galactic globular clusters (GC) - less than 150 (Harris 
\cite{harris96}) - were discovered mostly through optical searches, 
that are obviously biased against highly obscured GCs. Since the 
Galaxy is estimated to have $160\pm20$ GCs (Harris \cite{harris91}), 
a certain number of them may still be hidden behind the Galactic disk. 
The Two Micron All Sky Survey (2MASS) offers an opportunity to carry 
a search for the missing GCs, and recently Hurt et al. (\cite{hurt99}, 
\cite{hurt00}) reported a serendipitous discovery of two new GCs: 
2MASS GC01 and 2MASS GC02 (GC01 and GC02 hereafter). 
Their estimates suggest $\rm A_V=18-20$ mag, rendering these clusters 
unobservable in the optical wavebands. A summary of the available data 
for the new clusters is given in Table~\ref{tbl-1} (Hurt et al. 
\cite{hurt00}). 

%--------------------------------------------------------------------
\begin{table}[t]
\begin{center}
\caption{Basic cluster data}
\label{tbl-1}
\begin{tabular}{cccc} \hline
\multicolumn{1}{c}{Name}&
\multicolumn{1}{c}{R.A. Dec}&
\multicolumn{1}{c}{Diameter}&
\multicolumn{1}{c}{Core Magn} \\
\multicolumn{1}{c}{}&
\multicolumn{1}{c}{(2000)}&
\multicolumn{1}{c}{(arcmin)}&
\multicolumn{1}{c}{$\rm K_s$} \\
\hline
\object{GC01}&18:08:21.81 $-$19:49:47&$3.3\pm0.2$&$7.2\pm0.4$ \\
\object{GC02}&18:09:36.50 $-$20:46:44&$1.9\pm0.2$&$5.5\pm0.2$ \\
\hline
\end{tabular}
\end{center}
\end{table}
%--------------------------------------------------------------------

Situated in the general direction of the Galactic center, the new GCs 
are probably members of the disc system of GCs, which provides a 
valuable test case for understanding of the evolution of metal rich 
stars. 

Deep imaging in the near-infrared (NIR) is the only feasible way to 
study the new clusters. Glass \& Feast (\cite{glass73}), and Davidge 
\& Simons (\cite{davidge91}), among others, have demonstrated the 
usefulness of the NIR photometry of GCs for reddening and 
distance determinations. Recently, Kuchinski, Frogel \& Terndrup 
(\cite{kuchinski95}), Ferraro et al. (\cite{ferraro00}) and Ivanov 
et al. (\cite{ivanov00}) demonstrated photometric techniques to 
estimate the metal abundance [Fe/H] of GCs using the slope of the 
red giant branch (RGB) in the $\rm K_s$ vs. $\rm J-K_s$ diagram, 
with accuracy of $0.1-0.15$ dex. 

The development of the NIR methods and the extremely high reddening 
towards \object{GC01} and \object{GC02} prompted us to carry out 
the first deep NIR $\rm JK_s$ photometry of the two clusters. We 
present here preliminary estimates on their distance and reddening. 
A more detailed analysis and additional data will be reported in a 
subsequent paper. 

\section{Observations and Data Reduction}
The clusters \object{GC01} and \object{GC02} were observed at the 
ESO-NTT with SOFI in different occasions during July 2000. We used 
the Large-Field setup with a plate scale of 0.292 arcsec 
$\rm pixel^{-1}$, and a field of view 5x5 arcmin that allows to 
cover the entire clusters. The total integration time was 15, 10 
and 15 minutes for \object{GC01}, and 25, 15 and 15 for \object{GC02}, 
respectively in J, H and Ks. These observations were taken under 
non-photometric conditions. We also acquired 1 minute calibration 
images under photometric conditions. The seeing ranged between 0.6 
and 0.8 arcsec. A 3-color composites of the two clusters are shown 
in Figs.~\ref{fig1} and \ref{fig2}.

%--------------------------------------------------------------------
\begin{figure}
%  \resizebox{\hsoze}{!}{includegraphics{ivanovFig1.eps}}
\epsfxsize 8.0cm 
% for color
%%%%%%\centerline{\epsfbox{ivanovFig1.eps}}
\vspace{7.0cm}
% for gray scale
%\centerline{\epsfbox{ivanovFig1bw.eps}}}
  \caption{Near infrared 3-color composite image of \object{GC01}. 
  The $\rm J$, 
  $\rm H,$ and $K_s$ bands are mapped onto blue, green and red, 
  respectively. The image is 4.98 arcmin on the side. North is up, 
  and East is to to the left.}
  \label{fig1}
\end{figure}
%--------------------------------------------------------------------

The observational strategy consisted in alternating images on the 
objects and on the nearby sky (ON-OFF technique) to account for the 
sky background variations. Due to the high brightness of a few stars 
in the field, the images were taken with a short integration time, 
typically 1.2 sec, and were coadded at each position. The total 
integration time at each position was 60 sec. We introduced a random 
shift of $\sim 5-10$ arcsec between each object frame to improve 
the bad pixel and cosmic ray correction. The sky frames were taken 
at $\sim10$ arcmin from the clusters.

%--------------------------------------------------------------------
\begin{figure}
\epsfxsize 8.0cm 
%%%%%%%%%\centerline{\epsfbox{ivanovFig2.eps}}
\vspace{7.0cm}
  \caption{Near infrared 3-color composite image of \object{GC02}. 
  The size  and orientation are the same as in Fig.~\ref{fig1}.}
  \label{fig2}
\end{figure}
%--------------------------------------------------------------------

The data reduction included flat-fielding, and sky subtraction using 
IRAF\footnote{IRAF is distributed by the National Optical Astronomy 
Observatories, which are operated by the Association of Universities 
for Research in Astronomy, Inc., under cooperative agreement with 
the National Science Foundation.}. The images were shifted to a common 
position with fractional shifts and linear interpolation, and median 
combined together to produce the final image. The photometric 
calibration was performed using 2MASS stars $(40-60,$ depending on the 
band) in the field of \object{GC01} as photometric standards, with 
typical $\rm r.m.s.=0.05-0.07$ mag. 
The stellar photometry of the final combined images was carried out 
using DAOPHOT II (Stetson\cite{stetson93}). In this letter we consider 
only stars with DAOPHOT errors less than 0.15 mag. The median 
averaged $\rm 1\sigma$ errors over 1 mag bins are shown in Fig. 3. 
There is also an additional observational error of $\sim0.01$ mag 
due to the sky background variations. The effects of the crowding 
will be studied in the main paper by adding artificial stars in 
the field, and than measuring their brightness. 

The brightest stars $\rm(K_s\leq10$ mag) in our photometry are affected 
by the non-linearity of the array. For the purposes of this letter we 
will use only the fainter stars. A more comprehensive analysis and a 
non-linearity correction will be presented in a future paper.

\section{Analysis\label{tab3expl}}

\begin{figure}
%  \resizebox{\hsoze}{!}{includegraphics{ivanovFig3.eps}}
\epsfxsize 7.9cm 
\centerline{\epsfbox{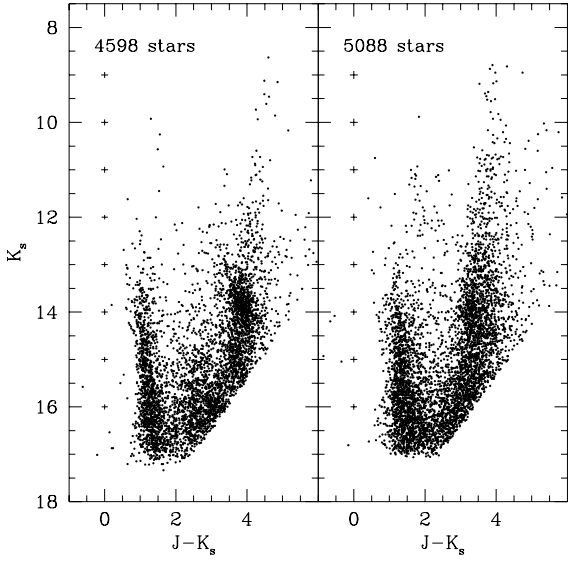}}
  \caption{The $\rm J-K_s$ vs. $\rm K_s$ color-magnitude diagrams of 
  all stars in the fields of \object{GC01} (left panel) and \object{GC02} 
  (right panel). The numbers of stars are indicated. DAOPHOT errors, 
  (median averaged over 1 mag bins) are shown on the left.}
  \label{fig3}
\end{figure}

The $\rm J-K_s$ vs. $\rm K_s$ color-magnitude diagrams (CMD) of the 
two clusters are shown in Fig.~\ref{fig3}. The most prominent 
features are the RGBs at $\rm J-K_s\sim3-4.5$ mag, and the main 
sequences (MS) of the field stars at $\rm J-K_s<2$ mag. They are 
both widened by the photometric uncertainties, and in the case of 
MS's - by distance and reddening variations. 

\begin{figure}
%  \resizebox{\hsoze}{!}{includegraphics{ivanovFig4.eps}}
\epsfxsize 7.9cm 
\centerline{\epsfbox{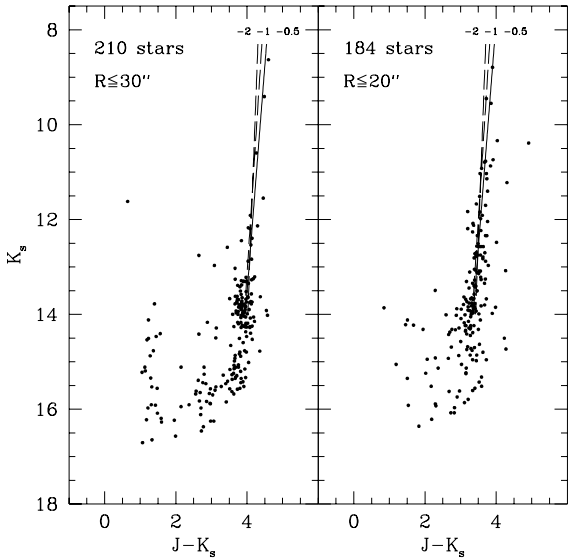}}
  \caption{The $\rm J-K_s$ vs. $\rm K_s$ color-magnitude diagrams of 
  the central $30\arcsec$ and $20\arcsec$ of \object{GC01} (left panel) 
  and \object{GC02} (right panel), respectively. The numbers of stars 
  are indicated. Fiducial lines, corresponding to the RGB slopes for 
  [Fe/H]$=-$2, $-$1, and $-$0.5 (from left to right) are shown.}
  \label{fig4}
\end{figure}

To eliminate the field contamination at least partially, we 
constructed the same CMDs for stars within 30 and 20 arcsec from the 
cluster centers, respectively for \object{GC01} and \object{GC02} 
(Fig.~\ref{fig4}). Although some blue disk stars are still present, 
the plots are dominated by cluster members. The red clumps 
are visible, at $\rm K_s=13.5-14.5$ mag. Since the RGB bump 
stars are not affected by the non-linearity, we can use them to 
deduce the cluster distances. 

The position of the RGB on the $\rm J-K_s$ vs. $\rm K_s$ CMD contains 
information about the metal abundance of the clusters. Fiducial lines, 
that represent slopes of RGBs for [Fe/H]$=-$0.5, $-$1, and $-$2, 
according to the calibration of Ivanov et al. (\cite{ivanov00}), are 
shown in Fig.~\ref{fig4}. Although the statistics is small, the plot is 
consistent with \object{GC02} being metal rich. Given the location of 
the GCs in the Galaxy, we will assume throughout the paper that both \object{GC01} and \object{GC02} have metallicity of [Fe/H]$=-$0.5. 

%--------------------------------------------------------------------
%--------------------------------------------------------------------
%--------------------------------------------------------------------
%--------------------------------------------------------------------

\begin{figure}
%  \resizebox{\hsoze}{!}{includegraphics{ivanovFig5.eps}}
\epsfxsize 7.9cm 
\centerline{\epsfbox{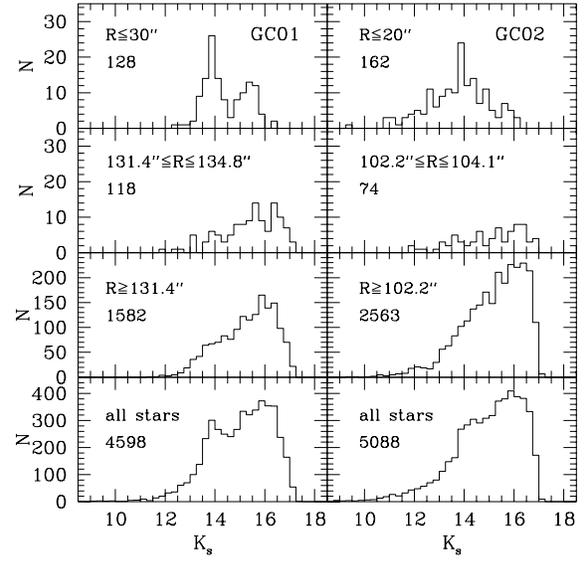}}
  \caption{Luminosity functions. \object{GC01} is on the left hand side, 
  and \object{GC02} 
  is on the right hand side. The top panels show the LFs in the central 
  regions. The second panels show the LFs for regions with the same 
  areas but outside of the cluster limits. The third panels show the 
  LFs for all non-cluster stars, and the bottom panels show the LFs for 
  the entire fields. The limits of the selected regions in terms of 
  distance from the cluster centers R, and the number of stars, included 
  in each LF are indicated on each panel.}
  \label{fig5}
\end{figure}

The RGB bump is more obvious on the $\rm K_s-band$ luminosity function 
(LF) plots (Fig.~\ref{fig5}). Again, we selected only stars near the 
cluster centers (top panels), to minimize the effect of the filed star 
contamination. For comparison purpose we also constructed the LFs for 
identical areas (second panels), well outside the cluster limits of 
$\rm(R_{cl}=99\pm6$ and $\rm57\pm6$ arcsec for \object{GC01} and 
\object{GC02}, Table~\ref{tbl-1}). Although the field star numbers are 
large, they are dominated by the faintest stars, which are underrepresented 
in the cluster LFs. The severe crowding in the clusters leads to shallower 
cut-offs in the cluster LFs. At the positions of the RGB bumps the field 
stars account for only about 15\% of the stars, and there is no trace of 
peaks. For completeness we plot the LFs of all non-cluster stars (third 
panels), and they also do not show detectable peaks at the position if 
the RGB bumps. Such traces are present in the LFs of the entire fields 
(bottom panels). 

The first step in our analysis was to estimate accurately the position 
of the bump. Following the example of Alves (\cite{alves00}) we fitted 
Gaussian to the $\rm K_s-band$ LFs in the regions of the bumps, 
obtaining $\rm K_s=13.86\pm0.30$ and $\rm K_s=13.99\pm0.25$ mag, 
respectively for \object{GC01} and \object{GC02}. The intrinsic spread 
of the stellar luminosities in the RGB bump accounts for the large 
uncertainties, in comparison with the photometric errors. 

%--------------------------------------------------------------------
%--------------------------------------------------------------------

Next, we applied the calibration of the absolute $\rm K-band$ magnitude 
of the 
RGB bump $\rm M_K^b$ as a function of the cluster metallicity, by Ferraro 
et al. (\cite{ferraro00}). They used the metallicity scale of Carretta \& 
Gratton (\cite{carretta97}). Although their $\rm K$ filter is different 
from our $\rm K_s$ filter, a comparison between the $\rm K$ and $\rm K_s$ 
measurements of red stars in Persson et al. (\cite{persson98}) yields 
differences of only $0.01-0.02$ mag, negligible for the purpose of this 
analysis. Henceforth we consider $\rm K$ and $\rm K_s$ magnitudes 
equivalent. 

The RGB bump brightness is rather sensitive to the metal abundance - 
according to Ferraro et al. (\cite{ferraro00}) it shifts from 
$\rm M_K^b=-0.92$ to $\rm M_K^b=-1.56$ for a metallicity change from 
[Fe/H]$=-$0.5 to $-1.$ We calculated $\rm(m-~M)_0+A_K=14.78\pm0.35$ and 
$14.91\pm0.31$ mag, respectively for \object{GC01} and \object{GC02}. 

To determine the reddening corrected distance moduli we used the color 
of the RGB. Ferraro et all. (\cite{ferraro00}) calibrated the RGB 
$\rm J-K$ color at a given $\rm M_K$ as a function of the metal content. 
Their faintest RGB level was $M_K=-3$ mag. Once we knew the brightness 
of the RGB bump, we could find the $M_K=-3$ mag level on our CMD, and 
calculated the $\rm (J-K)_{M_K=-3}$ color. We averaged the color of the 
stars over $\pm0.5$ mag interval, to ensure sufficient statistics. We 
estimated $\rm (J-K_s)_{M_K=-3}=4.13\pm0.11$ and $3.51\pm0.18$ mag for 
\object{GC01} and \object{GC02}, respectively. The predicted intrinsic 
RGB color for [Fe/H]$=-$0.5 
was $\rm (J-K)_{0,M_K=-3}=0.58\pm0.02$ mag, and the color 
excesses were $\rm E(J-K_s)=3.55\pm0.22$ and $2.93\pm0.18$ mag for 
\object{GC01} 
and \object{GC02}. We used the reddening curve of Rieke \& Lebofski (\cite{rieke85}) 
to calculate the extinction: $\rm A_V=20.88\pm0.65$ and $17.24\pm1.23$, 
$\rm A_K=2.34\pm0.07$ and $1.93\pm0.14$ mag for \object{GC01} and 
\object{GC02}. Finally, 
we obtained the reddening corrected distance moduli 
$\rm(m-M)_0=12.44\pm0.36$ and $12.98\pm0.41$ mag, and distances 
$\rm D=3.1\pm0.5$ and $3.9\pm0.6$ kpc for \object{GC01} and 
\object{GC02}, respectively. 
We performed the same estimates (Table~\ref{tbl-2}) for [Fe/H]$=-$1 
and $-$2, although the latter seems unlikely, given the cluster location. 

%--------------------------------------------------------------------
\begin{table}[t]
\begin{center}
\caption{Reddening and distance estimates. $1\sigma$ errors are given 
in brackets. See Sect.~\ref{tab3expl} for details.}
\label{tbl-2}

\begin{tabular}{lrr} \hline
\multicolumn{1}{c}{Parameter}&
\multicolumn{1}{c}{\object{GC01}}&
\multicolumn{1}{c}{\object{GC02}} \\
\hline
$\rm m_K^b$          &13.86(0.30)&13.99(0.25) \\
\hline
[Fe/H]               &\multicolumn{2}{c}{$-$0.5} \\ 
$\rm M_K^b$          &\multicolumn{2}{c}{$-$0.92(0.19)} \\ 
$\rm (m-M)_0+A_K$    &14.78(0.35)&14.91(0.31) \\
$\rm (J-K)_{M_K=-3}$ & 4.13(0.11)& 3.51(0.18) \\
$\rm (J-K)_0$        &\multicolumn{2}{c}{0.58(0.02)} \\
$\rm E(J-K)$         & 3.55(0.22)& 2.93(0.18) \\
$\rm A_V$            &20.88(0.65)&17.24(1.23) \\
$\rm A_K$            & 2.34(0.07)& 1.93(0.14) \\
$\rm (m-M)_0$        &12.44(0.36)&12.98(0.41) \\
D(kpc)               &  3.1(0.5) & 3.9(0.6)   \\
\hline
[Fe/H]               &\multicolumn{2}{c}{$-$1.0} \\ 
$\rm M_K^b$          &\multicolumn{2}{c}{$-$1.56(0.27)} \\ 
$\rm (m-M)_0+A_K$    &15.42(0.40)&15.55(0.37) \\
$\rm (J-K)_{M_K=-3}$ & 4.08(0.13)& 3.50(0.14) \\
$\rm (J-K)_0$        &\multicolumn{2}{c}{0.69(0.01)} \\
$\rm E(J-K)$         & 3.39(0.13)& 2.81(0.14) \\
$\rm A_V$            &19.94(0.76)&16.53(0.82) \\
$\rm A_K$            & 2.23(0.09)& 1.85(0.14) \\
$\rm (m-M)_0$        &13.19(0.41)&13.70(0.38) \\
D(kpc)               &  4.3(0.8) & 5.5(1.0)   \\
\hline
[Fe/H]               &\multicolumn{2}{c}{$-$2.0} \\ 
$\rm M_K^b$          &\multicolumn{2}{c}{$-$2.22(0.40)} \\ 
$\rm (m-M)_0+A_K$    &13.08(0.50)&12.21(0.47) \\
$\rm (J-K)_{M_K=-3}$ & 3.94(0.16)& 2.70(0.16) \\
$\rm (J-K)_0$        &\multicolumn{2}{c}{0.75(0.01)} \\
$\rm E(J-K)$         & 3.19(0.16)& 2.70(0.16) \\
$\rm A_V$            &18.76(0.94)&15.88(0.94) \\
$\rm A_K$            & 2.10(0.11)& 1.78(0.11) \\
$\rm (m-M)_0$        &13.98(0.51)&14.43(0.48) \\
D(kpc)               &  6.3(1.5) & 7.7(1.7)   \\
\hline
\end{tabular}
\end{center}
\end{table}
%--------------------------------------------------------------------

\section{Conclusions}
We carried out the first deep NIR photometry of \object{GC01} and 
\object{GC02} - new 
Galactic GCs, discovered by the 2MASS. The data suggest that they 
are metal rich GCs, consistent with their location in the disk. We 
detected the RGBs and the RGB bumps and estimated their distance 
and reddening using the brightness of RGB bumps and the color of 
the RGB. We found $\rm D=3.1\pm0.5$, $3.9\pm0.6$ kpc, and 
$\rm A_V=~20.9\pm0.7$, $17.2\pm1.2$, for \object{GC01} and 
\object{GC02} respectively. Here, we assumed [Fe/H]$=-$0.5, typical 
for disk clusters, but a different metal abundance could change 
our results within $30-35$\%.

Our data suggest extinction similar to the $\rm A_V=20\pm2$ and 
$18\pm2$ mag for \object{GC01} and \object{GC02}, obtained by Hurt 
et al. (\cite{hurt00}). A better knowledge of the metal abundance 
is crucial for an accurate distance and reddening determination 
with this technique. We will report a more sophisticated Monte-Carlo 
foreground removal method, metallicity estimates, and deeper CMDs 
and color-color diagrams in a subsequent paper. 

The authors are grateful to the referee Dr. R. Hurt for the helpful 
comments.

\end{document}